\documentclass[preprint,showpacs,nofootinbib]{revtex4}
\usepackage{mathrsfs}
\usepackage{float,epsfig}
\usepackage{dcolumn}% Align table columns on decimal point
\usepackage{bm}% bold math
\usepackage{graphicx}% Include figure files
\usepackage{bm}% bold math
\usepackage{subfigure}
\usepackage{amsmath,amssymb,amsthm}
\usepackage[colorlinks=true,linkcolor=blue]{hyperref}
\begin{document}
\title{\bf Phase transition and thermodynamical geometry for Schwarzschild AdS black hole in $AdS_5\times{S^5}$ spacetime}
\author{Jia-Lin Zhang$^{1,2}$~\footnote{Email: jlzhang1981@gmail.com},  Rong-Gen Cai$^{1,2}$~\footnote{Email: cairg@itp.ac.cn} and Hongwei Yu$^{1,3}$~\footnote{Email: hwyu@hunnu.edu.cn}}
\affiliation{$^1$ Department of Physics and Key Laboratory of Low Dimensional Quantum Structures and Quantum Control of Ministry of Education, Hunan Normal University, Changsha 410081, China\\
$^2$ State Key Laboratory of Theoretical Physics, Institute of Theoretical Physics,
Chinese Academy of Sciences, Beijing 100190, China \\
$^3$ Center for Nonlinear Science and Department of Physics, Ningbo University, Ningbo 315211, China}
%\footnote{Corresponding author}

\begin{abstract}

We study the thermodynamics and thermodynamic geometry of a
five-dimensional Schwarzschild AdS black hole in $AdS_5\times{S^5}$
spacetime by treating the cosmological constant as the number of
colors in the boundary gauge theory and its conjugate quantity as
the associated chemical potential. It is found that the chemical
potential is always negative in the stable branch of black hole
thermodynamics and it has a chance to be positive, but appears in
the unstable branch. We calculate the scalar curvatures of the
thermodynamical Weinhold metric, Ruppeiner metric and Quevedo
metric, respectively and we find that the scalar curvature in the
Weinhold metric is always vanishing, while in the Ruppeiner metric
the divergence of the scalar curvature is related to the divergence of
the  heat capacity with fixed chemical potential, and in the Quevedo
metric the divergence of the scalar curvature is related to the
divergence of the  heat capacity with fixed number of colors and to the
vanishing of the heat capacity with fixed chemical potential.

\end{abstract}
%%%%\pacs{12.20.Ds, 42.50.Ct,03.70.+k,04.70.Dy}
\pacs{04.60.-m, 04.70.Dy, 11.25.-w} \maketitle

\section{Introduction}

The thermodynamical properties of black holes in anti-de Sitter (AdS) space are quite different from
those of black holes in asymptotically flat or de Sitter  space. The main reason is that the AdS space acts as
a confined cavity so that black holes in AdS space can be thermodynamically stable. In particular, there exists a minimal
Hawking temperature for a Schwarzschild black hole in AdS space, below which there does not exist any black hole solution, instead a stable thermal gas solution exists.
For a given temperature above the minimal one, there exist two black hole solutions. The black hole with smaller horizon is thermodynamically unstable with a negative heat capacity, while the black
hole with larger horizon is thermodynamically stable with a positive heat capacity. And Hawking and Page find that a phase transition, named Hawking-Page
phase transition, will happen  between the stable large black hole and thermal gas in AdS
space~\cite{Hawking}. According to AdS/CFT correspondence, which says that there is
an equivalence between a weakly coupled gravitational theory in $d$-dimensional AdS
spacetime and a strongly coupled conformal field theory (CFT) in a $(d-1)$-dimensional
boundary of the AdS space~\cite{Gubser:1998bc,Maldacena:1997re,Witten:1998qj,Witten:1998qj2} (for a review, see~\cite{Aharony}),
thermodynamical properties of black holes in AdS space can be identified with those of dual strongly coupled CFT.
The Hawking-Page phase transition for black holes in AdS space is interpreted as the confinement/deconfinement
phase transition in gauge theory~\cite{Witten:1998qj2}.
Thus it becomes quite interesting to study thermodynamics and phase structure of black holes in AdS space. Indeed, in the past
few years there have been a lot of works on thermodynamics and phase transition for black holes in AdS space.

In ordinary thermodynamic systems, a divergence of heat capacity is
usually associated with a second order phase transition. For
Kerr-Newmann black holes in Einstein-Maxwell theory, some heat
capacities diverge at some black hole parameters. Based on this,
Davies argued that some second order phase transitions will happen
in Kerr-Newmann black holes~\cite{Davies}. For a
Reissner-Nordstr\"om AdS (RN-AdS) black hole, such a phase
transition was studied in some details in~\cite{Myers,YuTian2012}.
In a canonical ensemble with a fixed charge, it was found that there
exists a phase transition between small and large black holes. This
phase transition behaves very like the gas/liquid phase transition
in a Van der Waals system~\cite{Myers,rgc1}. However, an
identification between RN-AdS black hole and Van der Waals system
was recently realized in~\cite{Kubiznak}, where the negative
cosmological constant plays the role as pressure, while its
conjugate acts as thermodynamic volume of the black hole in the
so-called extended phase space~\cite{Dolan1,Dolan2} (for a recent
review, see \cite{Dolan3}). Recently, thermodynamics and phase
transition in the extended phase space for black holes in AdS space
have been extensively studied in the literature (for an incomplete list see \cite{others}.)

In the framework of AdS/CFT correspondence, the negative
cosmological constant is related to the degrees of freedom of dual
CFT. Thus it is an interesting question as to whether the
interpretation of the cosmological constant as pressure is
applicable to the boundary CFT. Very recently, it was argued that it
is more suitable to view the cosmological constant as the number of
colors in gauge field and its conjugate as associated chemical
potential~\cite{Dolan2014,Johnson,Kastor:2014dra}. This
interpretation was examined in the case of $AdS_5\times S^5$, for
${\cal N}$=4 supersymmetric
 Yang-Mills theory at large $N$ in \cite{Dolan2014}. The chemical potential conjugate to the number of colors, is calculated.
 It is found that the chemical potential in the high temperature phase of the Yang-Mills theory is negative and decreases as temperature increases. For spherical black holes in the bulk the chemical potential approaches zero as the temperature is lowered below the Hawking-Page temperature and changes its sign at a temperature near the temperature at which the heat capacity diverges.

On the other hand, applying the geometrical ideas to ordinary
thermodynamical systems  gives us an alternative way to study phase
transition in those systems. Weinhold~\cite{Weinhold}
first introduced a sort of metric defined as the second derivatives
of internal energy with respect to entropy and other extensive
quantities of a thermodynamic system. Soon later, based on the
fluctuation theory of equilibrium thermodynamics,
Ruppeiner~\cite{Ruppeiner-2} introduced another metric which is
defined as the minus second derivatives of
entropy  with respect to the internal energy and other extensive
quantities of a thermodynamic system. It was argued that the scalar curvature of the Ruppeiner metric
can reveal the micro interaction behind the thermodynamic system and its divergence  is
 related to some phase transition in the thermodynamical system~\cite{Rupreview}. In addition, it was shown that the
Weinhold metric is conformal to the Ruppeiner metric~\cite{Salamon}.
However, both of the Weinhold metric and Ruppeiner metric are  not
invariant under Legendre transformation and sometimes contradictory
results will be produced~\cite{Berry,Mrugala}. In order to solve
this puzzle, Quevedo {\it et al.}
\cite{Quevedo-1,Quevedo-2,Quevedo-3,Quevedo-4} proposed a method to
obtain a new formulism of Geometrothermodynamics whose metric is
Legendre invariant in the space of equilibrium states. To the best
of our knowledge, applying the thermodynamical geometry to black
hole thermodynamics was initiated in \cite{Ferrara:1997tw}, there it
was found that the Weinhold metric is proportional to the metric on
the moduli space for supersymmetric extremal black holes, whose
Hawking temperature is zero, and the Ruppeiner metric governing
fluctuations naively diverges, which is consistent with the argument
that near the extremal limit, the thermodynamical description breaks
down. Applying the thermodynamical geometry approach to phase
transition of black holes was followed in \cite{CC,Aman:2003ug}, and
for more relevant references see the recent review~\cite{Rup2013}
and references therein. In particular, Ref. \cite{Mansoori:2013pna}
has investigated the relation between the divergence of the scalar
curvature of thermodynamical geometry in different ensembles and the
singularity of heat capacities.

In this paper, we will study thermodynamics and thermodynamical
geometry for a five-dimensional Schwarzschild AdS black hole in
$AdS_5\times S^5$ by viewing the number of colors as a
thermodynamical variable from the view of point of dual CFT.  In
next section, we will review some basic thermodynamic properties  of a
black hole in $AdS_5\times{S^5}$ spacetime  by
treating the cosmological constant in the bulk  as the number of
colors~\cite{Dolan2014}. In Sec.~\ref{III} we will calculate the thermodynamical
curvatures of the Weinhold metric, Ruppeiner metric and Quevedo
metric, respectively, for the thermodynamical system, in order to
see the relations between the thermodynamical curvature and phase
transition. Note that such calculations can not be done if one views
the cosmological constant as a true constant, or even in the case of
the extended phase space in the sense~\cite{Dolan1,Dolan2}, because
in the latter case, the heat capacity $C_V$  always vanishes. We end
the paper with conclusions in Sec.~\ref{IV}.

\section{Thermodynamics of Schwarzschild AdS black hole in $AdS_5\times S^5$}
\label{II}

In this section, we will review the main  results obtained by Dolan
in Ref.~\cite{Dolan2014}.  In $AdS_5\times{S^5}$ spacetime, the
line element for a five-dimensional  Schwarzschild AdS black hole
reads~\cite{Myers}
\begin{equation}
\label{Sch}
ds^2=-fdt^2+\frac{1}{f}dr^2+r^2h_{ij}dx^idx^j+L^2d\Omega_{5}^2,
\end{equation}
where $d\Omega_{5}^2$ is the metric of a five-dimensional sphere
with unit radius, $h_{ij} dx^idx^j$ is the line element of a
three-dimensional Einstein space $\Sigma_3$ with constant curvature
$6k$, and the metric function $f$ is given by
\begin{equation}\label{fr}
f=k-\frac{m}{{r^2}}+\frac{r^2}{L^2}\;,
\end{equation}
 where $L$ is the $AdS$ radius and $m$ is an integration constant. The cosmological constant
is $\Lambda=-6/L^2$. Without loss of generality, one can take the scalar curvature parameter $k$ of the
three-dimensional space $\Sigma_3$ as $k=1$, $0$, or $-1$,
respectively.  The ten-dimensional spacetime (\ref{Sch}) can be viewed as the near horizon geometry of $N$ coincident
$D3$-branes in type IIB supergravity. In that case, the AdS radius $L$ has a relation to the number $N$ of D3-branes~\cite{Maldacena:1997re}
\begin{equation}\label{n1}
L^4=\frac{\sqrt{2}N\ell_p^4}{\pi^2},
\end{equation}
where $\ell_p$ is the ten-dimensional Planck length. According to AdS/CFT correspondence, the spacetime (\ref{Sch}) can be
regarded as the gravity dual to ${\cal N}$=4 supersymmetric Yang-Mills theory. Then  $N$ is nothing, but the rank of the gauge group of the supersymmetric $SU(N)$ Yang-Mills Theory. In the large $N$ limit, the number of degrees of freedom of the ${\cal N}$=4 supersymmetric Yang-Mills theory is proportional to $N^2$ (in fact, it is that of $8N^2$ massless bosons and fermions in the weak coupling limit~\cite{Gubser}).

 The event horizon
$r_h$ of the black hole is determined by the equation $f=0$. Then
according to Eq.~(\ref{fr}), the mass of the black hole  can be
expressed as
\begin{equation}\label{m1}
M=\frac{ 3\omega_3 }{16 \pi G_5}m=\frac{3 \omega_3{r_h}^2}{16\pi
G_5L^2}(k L^2+r_h^2),
\end{equation}
where $\omega_3$ is the volume of $\Sigma_3$. Using the
Bekenstein-Hawking entropy formula of the black hole, we have
\begin{equation}
S=\frac{A}{4G_5}=\frac{\omega_3 r_h^3}{4G_5}\;.
\end{equation}
Note that $G_5=G_{10}/(\pi^3L^5)$ and $G_{10}=\ell_p^8$.
 %%%%%%%%%%%%%%
Therefore, the  mass of the black hole can be rewritten as a function of
$N$ and $S\;$
\begin{equation}\label{m2}
M(S,N)=\frac{3\tilde{m}_p}{4}\bigg[k\bigg(\frac{S}{\pi}\bigg)^{\frac{2}{3}}N^{\frac{5}{12}}+
\bigg(\frac{S}{\pi}\bigg)^{\frac{4}{3}}N^{-\frac{11}{12}}\bigg]\;,
\end{equation}
where $\tilde{m}_p=\sqrt{\pi}\ell_p^7/(2^{1/8}G_{10})$ is associated
with the 10-dimensional Planck mass. According to the standard
thermodynamic relation $dM=TdS+{\mu}dN^2$, the temperature
can be obtained
\begin{equation}\label{temperature}
T=\frac{\partial{M}}{\partial{S}}\bigg|_N=\frac{\tilde{m}_p}{2\pi}\bigg[k\bigg(\frac{S}{\pi}\bigg)^{-\frac{1}{3}}N^{\frac{5}{12}}+
2\bigg(\frac{S}{\pi}\bigg)^{\frac{1}{3}}N^{-\frac{11}{12}}\bigg]\;,
\end{equation}
which is nothing but the Hawking temperature of the black hole.
 The chemical potential $\mu$ conjugate to the number of colors is
 \begin{equation}\label{chem}
 \mu=\frac{\partial{M}}{\partial{N^2}}\bigg|_S=\frac{\tilde{m}_p}{32}\bigg[5
 k\bigg(\frac{S}{\pi}\bigg)^{\frac{2}{3}}N^{-\frac{19}{12}}-
11\bigg(\frac{S}{\pi}\bigg)^{\frac{4}{3}}N^{-\frac{35}{12}}\bigg]\;,
 \end{equation}
 which is the measure of the energy cost to the system when one increases  the number of colors.
 %%%%%%%%%%%%%%%

The Gibbs free energy  can be calculated as
\begin{equation}
\label{gibbs}
G(T,N^2)=M-TS=\frac{\tilde{m}_p}{4}\bigg[
 k\bigg(\frac{S}{\pi}\bigg)^{\frac{2}{3}}N^{\frac{5}{12}}-
\bigg(\frac{S}{\pi}\bigg)^{\frac{4}{3}}N^{-\frac{11}{12}}\bigg]\;.
\end{equation}

For the cases of $k=0$ or $k=-1$, it is easy to see  from
Eq.~(\ref{temperature}) for a fixed $N^2$ that the Hawking
temperature increases monotonically with the entropy $S$. Besides, we see from Eq.~(\ref{chem})and Eq.~(\ref{gibbs}) that
when $k=0$ or $k=-1$, the chemical potential is always negative, no
phase transition happens. However, when $k=1$, the situation is
quite different.

\begin{figure}[htbp]
\centering
\includegraphics[scale=0.8]{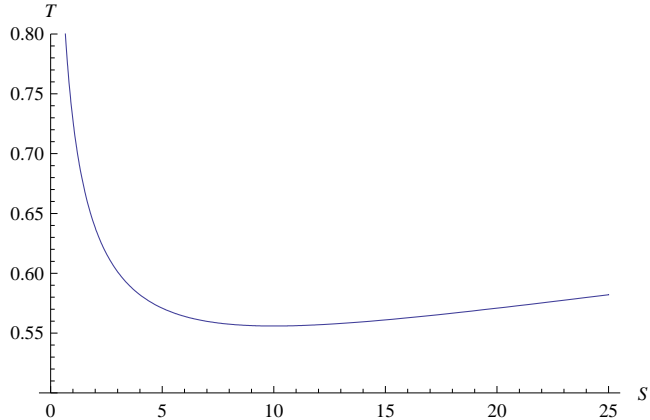}
\caption{The temperature with respect to entropy. Here we take
$\ell_p=1,\ k=1$ and $N=3$. The temperature arrives at the minimal
value when $S=S_1=N^2\pi/2^{3/2}\approx9.9965$.}\label{temp}
\end{figure}

In the case of $k=1$, the Hawking temperature is not a monotonic
function but has a minimum at
\begin{equation}
S_1=N^2\pi/2^{3/2}\;,
\end{equation}
or equivalently, at $r_h=L/\sqrt{2}$. We plot the behavior of
temperature with respect to entropy  in Fig.~\ref{temp}. The
corresponding minimal temperature is
$T_{\infty}=\sqrt{2}\tilde{m}_p/(\pi{N^{1/4}})= \sqrt{2}/(\pi L)$.
Namely under the minimal temperature there is no black hole
solution. Above the minimal temperature, there exist two branches,
as we will see shortly, the branch with small entropy (horizon
radius) is thermodynamically unstable, while the branch with large
entropy (horizon radius) is thermodynamically stable.

\begin{figure}[htbp]
\centering
\includegraphics[scale=0.8]{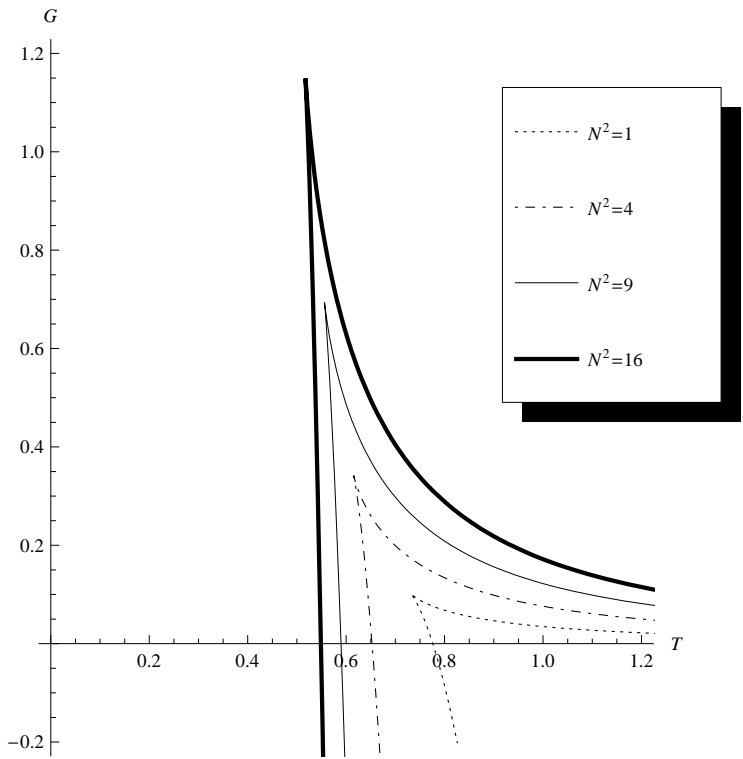}
\caption{The Gibbs free energy  as a function of the temperature for
various numbers of colors. Here we  take
 $\ell_p=1$ and $k=1$. The down branch Gibbs free energy  for a fixed $N$ changes its sign at the point $S=S_2=N^2\pi$,
 which corresponds to the Hawking-Page transition point. }\label{gt}
\end{figure}

One can see easily from the Gibbs free energy  Eq.~(\ref{gibbs}),
the Hawking-Page phase transition happens at $r_h=L$ with the phase transition
temperature $T_*= 3\tilde{m}_p/(2\pi{N^{1/4}})=3/(2\pi L)$, which is
larger than $T_{\infty}$. And the corresponding entropy at the
Hawking-Page transition is
\begin{equation}
S_2= N^2\pi\;.
\end{equation}
We can see that $S_2 >S_1$. In Fig.~\ref{gt}, we show the Gibbs free
energy with respect to the Hawking temperature $T$ for some fixed
$N$.

\begin{figure}[htbp]
\centering
\includegraphics[scale=0.8]{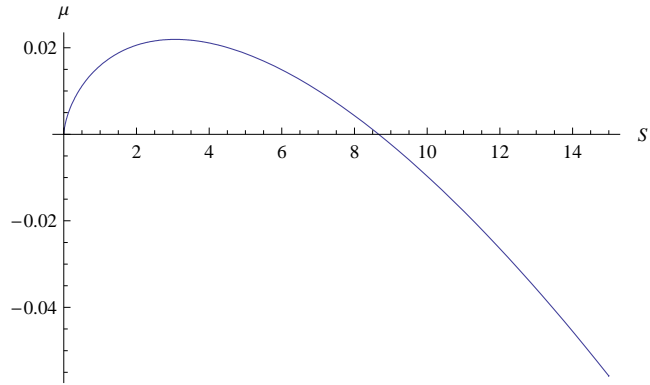}
\caption{The chemical potential as a function of entropy for a fixed
$N=3$. Here we take $\ell_p=1$ and $k=1$. The chemical potential
changes its sign at
$S=S_3=N^2\pi(5/11)^{3/2}\approx8.6648$.}\label{mus}
\end{figure}

\begin{figure}[htbp]
\centering
\includegraphics[scale=0.8]{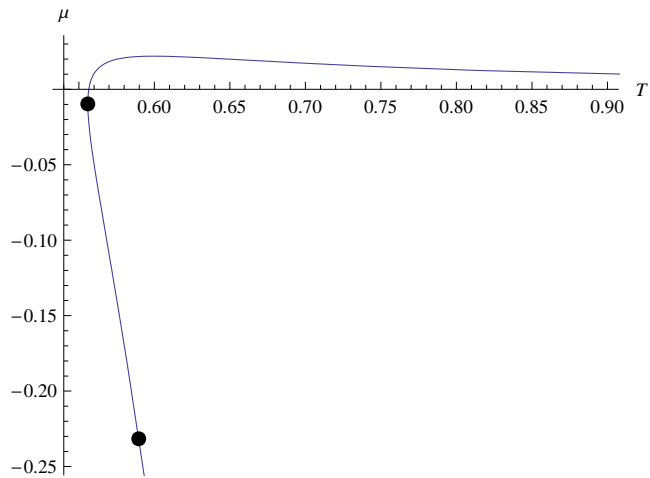}
\caption{The chemical potential as a function of temperature $T$ for
a fixed $N=3$. Here we take $\ell_p=1$ and $k=1$. Note that the
upper dot denotes the minimal temperature $T_{\infty}$ and the lower
dot denotes the Hawking-Page transition temperature $T_{*}\;.$  }\label{mut}
\end{figure}

\begin{figure}[htbp]
\centering
\includegraphics[scale=0.8]{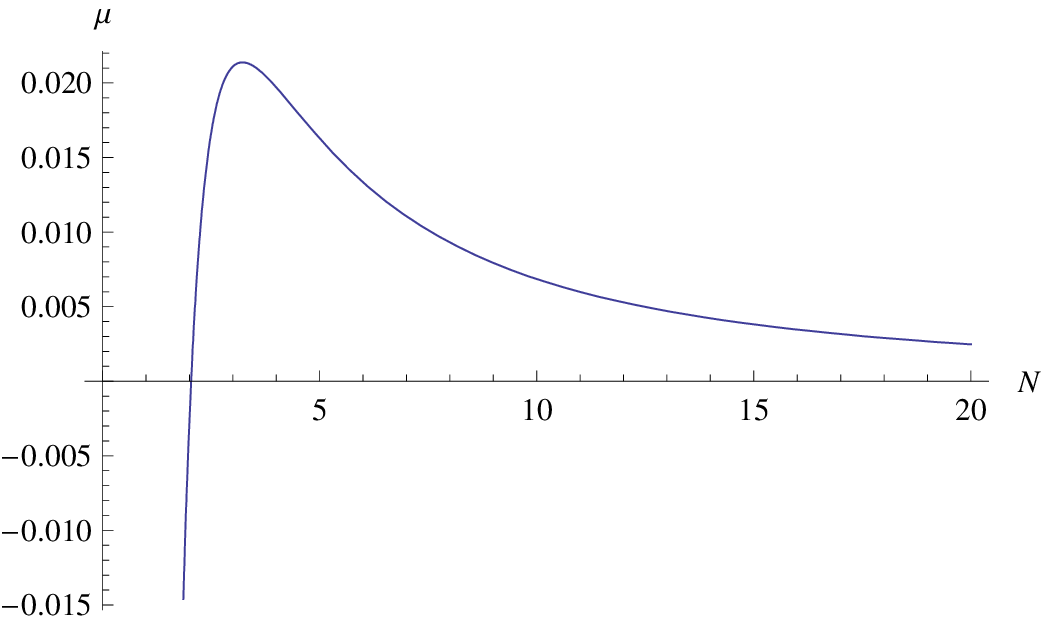}
\caption{The chemical potential  as a function of $N$ for a fixed
$S=4$. Here we take $\ell_p=1$ and $k=1$. The maximum of the
chemical potential corresponds to the point with
$S=S_5=N^2\pi{19}^{3/2}/{77}^{3/2}$, namely, $N\approx3.2230.$
}\label{mun}
\end{figure}

In Fig.~\ref{mus} we show the chemical potential as a function of
entropy $S$ for a fixed $N$. We see that the chemical potential is
positive when $S$ is small, while it changes to be negative when $S$
is large. The chemical potential changes its sign  at
\begin{equation}
 S_3=N^2\pi(5/11)^{3/2}\;.
 \end{equation}
We see that
 \begin{equation}
S_3 <S_1 <S_2.
 \end{equation}
%%%%%%%%%%%%%%%%%%%%%%%%%%%%%%%%%%%%%%%%
As we will see that the vanishing of the chemical potential appears
in the unstable branch. This implies that the vanishing of the
chemical potential does not make any sense from the point of view of
dual supersymmetric Yang-Mills theory.
 In Fig.~\ref{mut} we plot the chemical potential as a function of temperature $T$ for a fixed $N$, while in Fig.~\ref{mun} the
 chemical potential is plotted as a function of $N$ in the case with a fixed entropy  $S$.

  In the following section, we will study thermodynamical geometry of the
 Schwarzschild AdS black hole in the extended phase space by viewing
 the cosmological constant as the number of colors. We pay attention to the case of
 $k=1$, since the cases of $k=0$ and $k=-1$ are trivial.

%%%%%%%%%%%%%%%%%%%%%%%%%%%%%%%%%%%%%%%%%%%%%%%
\section{Thermodynamical geometry of the Schwarzschild AdS black hole}
\label{III}

\begin{figure}[htbp]
\centering
\includegraphics[scale=0.8]{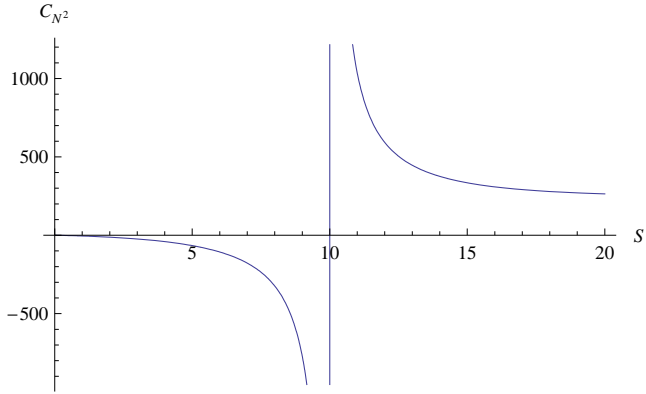}
\caption{The  heat capacity  in the case with a fixed $N=3$ as a
function of entropy  $S$. Here we take $k=1$ and $\ell_p=1 $.  The
divergence corresponds to the point
$S=S_1=N^2\pi/2^{3/2}\approx9.9965.$}\label{cneps}
\end{figure}

When the corresponding number of colors $N^2$ is kept fixed, this corresponds to the case in a
 canonical ensemble. In this case, the  heat capacity  for a fixed $N^2$ can be
obtained as
\begin{equation}\label{CN}
C_{N^2}=T\bigg(\frac{\partial{S}}{\partial{T}}\bigg)_{N^2}=\frac{3S(
N^{4/3}\pi^{2/3}+2S^{2/3})}{2S^{2/3}-N^{4/3}\pi^{2/3}}\;.
\end{equation}
The  heat capacity  diverges at the point of $S_1=N^2\pi/2^{3/2}$
(i.e., $r_h=L/\sqrt{2}$) which just coincides with the point
corresponding to the minimal Hawking temperature for a fixed $N^2$.
When $S<S_1$, the heat capacity is negative, indicating the
thermodynamical instability, while it is positive as $S>S_1$.  We
show the behavior of $C_{N^2}$  as a function of $S$ in
Fig.~\ref{cneps}.

\begin{figure}[htbp]
\centering
\includegraphics[scale=1]{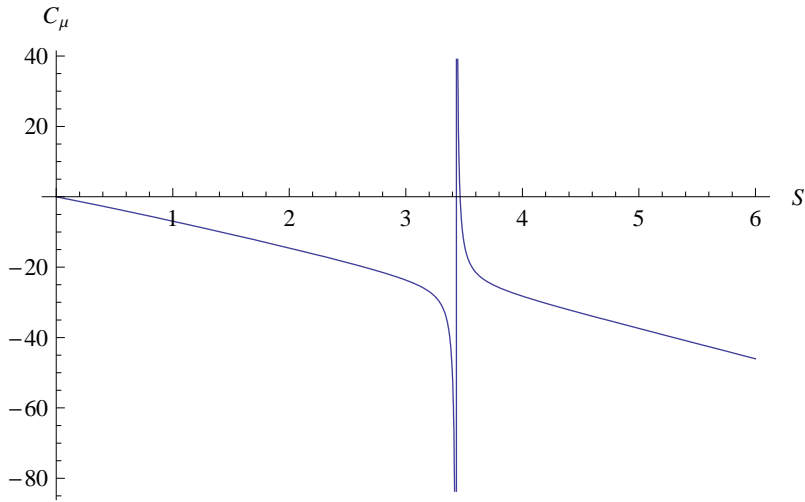}
\caption{The  heat capacity for a fixed $\mu$ vs  entropy  $S$ for
$N=3$,$k=1$ and  $\ell_p=1$.  The divergence corresponds to the
point
$S=S_4=5\sqrt{2}\pi{N^2}/{\sqrt{1665+67\sqrt{665}}}\approx3.4324$.
The heat capacity  vanishes  at a nontrivial point
$S=S_5=N^2\pi{19}^{3/2}/{77}^{3/2}\approx3.4657.$ Note that there is
a trivial zero  heat capacity  at $S=0$, which will not be
considered here. }\label{cmu}
\end{figure}

In the grand canonical ensemble with fixed chemical potential $\mu$,
corresponding  heat capacity can be obtained as
\begin{equation}
C_{\mu}=T\bigg(\frac{\partial{S}}{\partial{T}}\bigg)_{\mu}=\frac{-95N^{8/3}\pi^{4/3}S+195N^{4/3}\pi^{2/3}S^{5/3}+770S^{7/3}}{15
N^{8/3}\pi^{4/3}-45N^{4/3}\pi^{2/3}S^{2/3}-66S^{4/3}}\;.
\end{equation}
The  heat capacity is plotted in
Fig.~\ref{cmu}. We see that the  heat capacity diverges  at
\begin{equation}
S=S_4=\frac{5\sqrt{2}\pi{N^2}}{\sqrt{1665+67\sqrt{665}}}
\end{equation}
which corresponds to the horizon radius $r_h\approx0.49515L\;.$
Clearly $S_4 <S_1$, namely the divergence happens in the small black
hole branch. There exists only a very limited region with a positive
 heat capacity between $ S_4 < S<S_5$, where
 \begin{equation}
S_5=N^2\pi\bigg(\frac{19}{77}\bigg)^{3/2}\;,
\end{equation}
namely, $r_h\approx0.49674L$, which has a vanishing heat capacity.
Note that $S_5$ is also less than $S_1$. This is quite different
from the classical gas with negative chemical potential. When the
chemical potential approaches zero and becomes positive, quantum
effects should come into playing some role~\cite{Dolan2014}.
%%%%%%%%%%%%%%

%%%%%%%%%%%%%%%%%%%%%%%%%%%%%%%

Now we turn to the thermodynamical geometry of the black hole and
want to see whether the thermodynamical curvature can reveal the singularity
of these two  heats capacities.  The Weinhold metric~\cite{Weinhold} is defined as the
second derivatives of internal energy with respect to entropy and
other extensive quantities in the energy representation, while the Ruppeiner metric \cite{Ruppeiner-2} is  related to the Weinhold metric by a conformal factor of temperature~\cite{Salamon}
\begin{equation}\label{conform}
ds^2_R=\frac{1}{T}ds^2_W\;.
\end{equation}
The Weinhold metric and  Ruppeiner metric, which are
 dependent on the choice of thermodynamic potentials, are not
Legendre invariant.

 Quevedo {\it et al.}~\cite{Quevedo-1,Quevedo-2,Quevedo-3,Quevedo-4} proposed a method
to obtain a thermodynamical metric from a Legendre invariant
thermodynamic potential. This method allows one to obtain a new
formulism of Geometrothermodynamics whose metric is Legendre
invariant in the space of equilibrium states. In what follows, we
will first briefly review the formulism of Geometrothermodynamics.
Define an (2n+1)-dimensional thermodynamic phase space $\mathcal{T}$
which can be described by the coordinates of $\{\phi,E^a,I^a\}\;,
a=1,\ldots,n\;,$ where $\phi$ denotes the thermodynamic potential,
$E^a$ and $I^a$ respectively represent the set of extensive
variables and the set of intensive variables. Then the fundamental
Gibbs 1-form can be defined on the space  $\mathcal{T}$ as
$\Theta=d\phi-\delta_{ab}I^adE^b$ with $\delta_{ab}={\rm
diag}(1,1,\ldots,1)\;.$ Under the assumption that
 $\mathcal{T}$ is differentiable and $\Theta$ satisfies the condition
 of $\Theta\wedge(d\Theta)^n\neq0\,,$ the pair $(\mathcal{T},\Theta)$ defines a contact manifold.
Considering G as a non-degenerate Riemannian
 metric  on the space  $\mathcal{T}$, especially, the geometric
 properties of metric G  do not depend on the choice of
 thermodynamic potential in its construction because of Legendre
invariance, then  the set $(\mathcal{T},\Theta,G)$ can define a
Riemannian contact manifold or the phase manifold. As a result, an
n-dimensional Riemannian submanifold $\varepsilon\subset\mathcal{T}$
can be defined as the space of thermodynamic equilibrium states
(equilibrium manifold) by a smooth map
$\varphi:\varepsilon\rightarrow\mathcal{T},$ i.e.,
$\varphi:(E^a)\mapsto(\phi,E^a,I^a)\;$  where the pullback of the
map should satisfy the condition $\varphi^*(\Theta)=0\;.$
Furthermore, Quevedo metric $g$ can be induced on the equilibrium
manifold $\varepsilon$ by  using $\varphi^*(G)\;.$ The
non-degenerate Riemannian
 metric G can be chosen as \cite{Quevedo-3}
 \begin{equation}
 G=(d\phi-\delta_{ab}I^adE^b)^2+(\delta_{ab}E^aI^b)(\eta_{cd}dE^cdI^d),\ \eta_{cd}={\rm diag}(-1,1,\ldots,1).
 \end{equation}
  Then Quevedo
metric reads
\begin{equation}\label{Gq}
g=\varphi^*(G)=\bigg(E^c\frac{\partial{\phi}}{\partial{E^c}}\bigg)
\bigg(\eta_{ab}\delta^{bc}\frac{\partial^2{\phi}}{\partial{E^c\partial{E^d}}}dE^adE^d\bigg)\;.
\end{equation}

Now we calculate the thermodynamical curvature for the Schwarzschild AdS black hole.  The Weinhold metric is given by
\begin{equation}\label{gW}
g^W=\Bigg(\begin{matrix}M_{SS} & M_{SN^2}\\
{M_{N^2S}}&M_{N^2N^2}\end{matrix}\Bigg)\;,
\end{equation}
where $\rho_{ij}$ stands for  $\partial^2{\rho}/\partial x^i\partial
x^j$, and $x^1=S$, $x^2=N^2$. The scalar curvature  of this metric
can be calculated directly. Substituting Eq.~(\ref{m2}) and
Eq.~(\ref{temperature}) into Eq.~(\ref{gW}), we can see that the scalar curvature of the
 Weinhold metric is always vanishing.

\begin{figure}[htbp]
\centering
\includegraphics[scale=1]{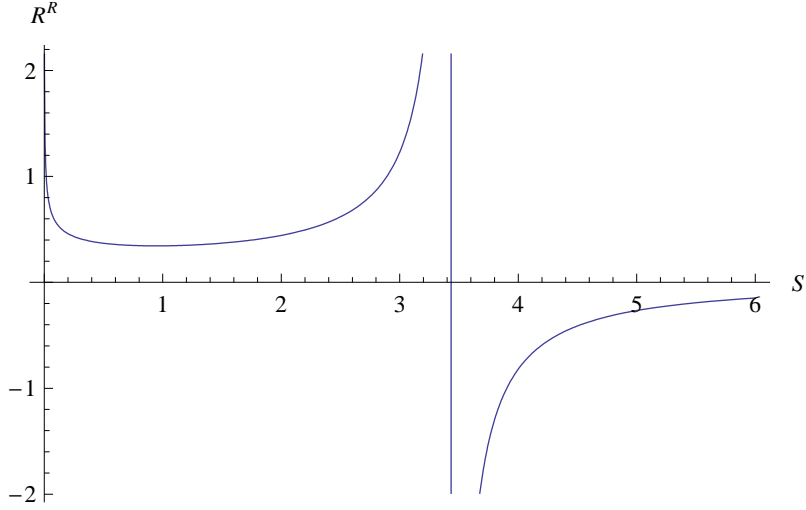} \caption{ The scalar
curvature vs  entropy  for the Ruppeiner metric case with $N=3$, $k=1$
and $\ell_p=1$.  The
 scalar curvatures diverge at
$S=S_4=5\sqrt{2}\pi{N^2}/\sqrt{1665+67\sqrt{665}}\approx3.4324.$
}\label{rrs}
\end{figure}

On the other hand, considering Eq.~(\ref{conform}), the Ruppeiner metric can be written
as
\begin{equation}\label{gR}
g^R=\frac{1}{T}\Bigg(\begin{matrix}M_{SS} & M_{SN^2}\\M_{N^2S}&
{M_{N^2N^2}}\end{matrix}\Bigg)\;,
\end{equation}
and the corresponding curvature of this metric is
\begin{equation}\label{RR}
R^R=\frac{7(5N^{8/3}\pi^{4/3}S^{-1/3}+2N^{4/3}\pi^{2/3}S^{1/3})}{15N^{4}\pi^{2}
-15N^{8/3}\pi^{4/3}S^{2/3}-156N^{4/3}\pi^{2/3}S^{4/3}-132S^2}\;.
\end{equation}
From Eq.~(\ref{RR}), we can conclude that the scalar curvatures of the
 Ruppeiner metric possess a positive
singularity at
$S=S_4=5\sqrt{2}\pi{N^2}/{\sqrt{1665+67\sqrt{665}}}\;$, i.e.,
$r_h\approx0.49515L\;$ (see Fig.~\ref{rrs}). This singularity just
coincides with the divergence of the heat  capacity $C_\mu$ for
fixed chemical potential (comparing Fig.~\ref{cmu} with
Fig.~\ref{rrs}). Therefore, we may conclude that the Ruppeiner metric
can reveal the  phase transition of the Schwarzschild AdS black hole
in $AdS_5\times{S^5}$ in grand canonical ensemble, while the Weinhold metric cannot here.

\begin{figure}[htbp]
\centering
\includegraphics[scale=1]{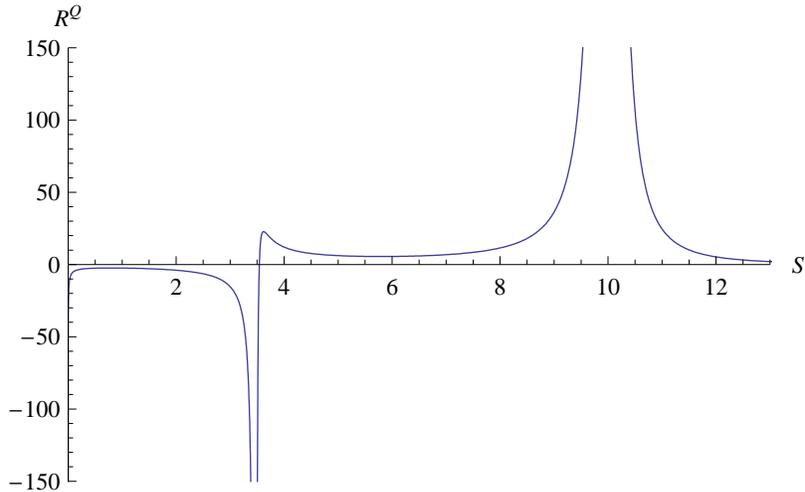}
\caption{Scalar curvature vs entropy $S$ for the Quevedo metric with
$N=3,\ell_p=1$ and $k=1$.  There exist two divergences at
$S=S_5=N^2\pi{19}^{3/2}/{77}^{3/2}\approx3.4657\;$ and
$S=S_1=N^2\pi/2^{3/2}\approx9.9965\;$, respectively. }\label{rrq}
\end{figure}

The  Quevedo metric reads
\begin{equation}
g^Q=(ST+N^2\mu)\Bigg(\begin{matrix}-M_{SS} &
0\\0&M_{N^2N^2}\end{matrix}\Bigg)\;.
\end{equation}
Calculating its scalar curvature gives
\begin{equation}
R^Q=A_1/B_1,
\end{equation}
where $A_1$ and $B_1$ are given by
\begin{eqnarray}
A_1&=&256N^{19/6}\pi^{10/3} \left(3982 S^2-1741 N^4 \pi
^2+2372N^{8/3}\pi^{4/3}S^{2/3}+17311\pi^{2/3}
N^{4/3}S^{4/3}\right)\;, \nonumber \\
 B_1 &=& 105 \tilde{m}_p^2 S^{2/3} \left(N^{4/3} \pi ^{2/3}+S^{2/3}\right)^3
  \left(19 N^{8/3} \pi ^{4/3}-115 N^{4/3} \pi ^{2/3} S^{2/3}+154
  S^{4/3}\right)^2\;.\nonumber
\end{eqnarray}
The scalar curvature is plotted in Fig.~\ref{rrq}, we see that there exist two divergent
points at
\begin{equation}
S_1=\frac{N^2\pi}{2\sqrt{2}}\ \ \ \ {\rm and}\ \
S_5=N^2\pi\bigg(\frac{19}{77}\bigg)^{3/2}\;,
\end{equation}
respectively. The first one just coincides with the divergent point of $C_{N^2}$, while the second one corresponds
to $C_{\mu}=0$.  This result is consistent with the recent study in \cite{Hendi,Suresh} that
 the divergences of scalar curvature for the Quevedo metric correspond to divergence or zero for  heat capacity.
 These results are meaningful to further understand the relation between phase transition and thermodynamical curvature.

\section{Conclusions}
\label{IV}

In this paper, we have studied thermodynamics of a Schwarzschild AdS black hole in $AdS_5 \times S^5$ spacetime in the
extended phase space where the cosmological constant is viewed as the number of colors in the dual supersymmetric Yang-Mills
theory. We calculated and discussed the chemical potential associated with the number of colors, and found that the
chemical potential is always negative in the stable branch of black hole thermodynamics. The chemical potential has a chance
to be positive, but it appears in the unstable branch.

The  heat capacities  with fixed number of colors $C_{N^2}$ and with fixed chemical potential $C_{\mu}$  have been calculated, respectively.
It is found that $C_{N^2}$ diverges at the minimal temperature of the black hole, while $C_{\mu}$ diverges at a smaller horizon radius.

In the extended phase space, we have a chance to study the
thermodynamical geometry associated with the Schwarzschild AdS black
hole. By calculating scalar curvatures of the Weinhold metric,
Ruppeiner metric and Quevedo metric, we see that in the  Weinhold
metric the scalar curvature is always zero, no singularity is found.
However, in the Ruppeiner metric  the scalar curvature diverges at same
divergent point of $C_{\mu}$, and in the Quevedo metric, the scalar
curvature diverges at the divergence of $C_{N^2}$, besides at the
point of $C_{\mu}=0$. These results indicate that the divergence of
thermodynamical curvature indeed is related to some divergence of
heat capacities, but the divergence of thermodynamical curvature may
be also related to the vanishing points of the thermodynamic
potential, temperature,  heat capacity,  etc.~\cite{Hendi,Suresh}.
This is helpful to further understand the relation between phase
transition and divergence of thermodynamical curvature. For a
further study of this relation, it should be of great interest to
discuss thermodynamics and thermodynamical curvature for other black
holes in AdS space in the extended phase space.

\begin{acknowledgments}
RGC thanks Hunan Normal University for warm hospitality extended to him during his visit.
 This work was supported in part by the National Natural Science
Foundation of China (No.11005038, No.10975168, No.11035008,
No.11375092 and No.11435006), the Hunan Provincial Natural Science
Foundation of China under Grant No. 11JJ7001 and the Program of
Excellent Talent of Hunan Normal University No. ET13102.
\end{acknowledgments}


\begin{thebibliography}{00}
\bibitem{Hawking}S. W. Hawking and D. N. Page, Commun.
Math.Phys.{\bf87}, 577 (1983).
\bibitem{Maldacena:1997re}
  J.~M.~Maldacena,
  %``The large N limit of superconformal field theories and supergravity,''
  Adv.\ Theor.\ Math.\ Phys.\  {\bf 2}, 231 (1998)
  [Int.\ J.\ Theor.\ Phys.\  {\bf 38}, 1113 (1999)].
  %[arXiv:hep-th/9711200].
  %%CITATION = IJTPB,38,1113;%%
%\cite{Gubser:1998bc}
\bibitem{Gubser:1998bc}
  S.~S.~Gubser, I.~R.~Klebanov and A.~M.~Polyakov,
  %``Gauge theory correlators from non-critical string theory,''
  Phys.\ Lett.\  B {\bf 428}, 105 (1998).
  %[arXiv:hep-th/9802109].
  %%CITATION = PHLTA,B428,105;%%
%\cite{Witten:1998qj}
\bibitem{Witten:1998qj}
  E.~Witten,
  %``Anti-de Sitter space and holography,''
  Adv.\ Theor.\ Math.\ Phys.\  {\bf 2}, 253 (1998).
  %[arXiv:hep-th/9802150].
  %%CITATION = 00203,2,253;%%

%\cite{Witten:1998qj2}
\bibitem{Witten:1998qj2}
E. Witten,
%¡°Anti-de Sitter space, thermal phase transition, and confinement in gauge theories,¡±
Adv. Theor. Math. Phys. {\bf 2}, 505 (1998).
%[arXiv:hep-th/9803131].

\bibitem{Aharony} O. Aharony, S.S. Guber, J. Maldacena, H. Ooguri and Y. Oz, Phys. Rep. {\bf 323}, 183 (2000).
\bibitem{Davies}P. C. W. Davies, Proc. Roy. Soc. Lond. A {\bf353}, 499 (1977); P. C. W. Davies, Rep. Prog. Phys. {\bf41}, 1313
(1977); P. C. W. Davies, Class. Quant. Grav. {\bf6}, 1909 (1989).

\bibitem{Myers}
  A.~Chamblin, R.~Emparan, C.~V.~Johnson and R.~C.~Myers,
  %``Charged AdS black holes and catastrophic holography,''
  Phys.\ Rev.\ D {\bf 60}, 064018 (1999)
  [hep-th/9902170].
  %%CITATION = HEP-TH/9902170;%%
  %451 citations counted in INSPIRE as of 27 Aug 2014

 \bibitem{YuTian2012}
 C.  Niu, Y. Tian, and X.  Wu,
%"Critical Phenomena and Thermodynamic Geometry of RN-AdS Black Holes,"
Phys. Rev.D {\bf85},  024017(2012) [arXiv:1104.3066 [hep-th]].


\bibitem{rgc1}J.~y.~Shen, R.~G.~Cai, B.~Wang and R.~K.~Su,
  %``Thermodynamic geometry and critical behavior of black holes,''
  Int.\ J.\ Mod.\ Phys.\ A {\bf 22}, 11 (2007)
  [gr-qc/0512035].
  %%CITATION = GR-QC/0512035;%%
  %73 citations counted in INSPIRE as of 27 Aug 2014


\bibitem{Kubiznak}D.~Kubiznak and R.~B.~Mann,
  %``P-V criticality of charged AdS black holes,''
  JHEP {\bf 1207}, 033 (2012)
  [arXiv:1205.0559 [hep-th]].
  %%CITATION = ARXIV:1205.0559;%%
  %61 citations counted in INSPIRE as of 27 Aug 2014

\bibitem{Dolan1}D.~Kastor, S.~Ray and J.~Traschen,
  %``Enthalpy and the Mechanics of AdS Black Holes,''
  Class.\ Quant.\ Grav.\  {\bf 26}, 195011 (2009)
  [arXiv:0904.2765 [hep-th]].
  %%CITATION = ARXIV:0904.2765;%%
  %74 citations counted in INSPIRE as of 27 Aug 2014

\bibitem{Dolan2}B.~P.~Dolan,
  %``Pressure and volume in the first law of black hole thermodynamics,''
  Class.\ Quant.\ Grav.\  {\bf 28}, 235017 (2011)
  [arXiv:1106.6260 [gr-qc]].
  %%CITATION = ARXIV:1106.6260;%%
  %56 citations counted in INSPIRE as of 27 Aug 2014

\bibitem{Dolan3}B.~P.~Dolan,
  %``Black holes and Boyle's law -- the thermodynamics of the cosmological constant,''
  arXiv:1408.4023 [gr-qc].
  %%CITATION = ARXIV:1408.4023;%%

\bibitem{others}S.~Gunasekaran, R.~B.~Mann and D.~Kubiznak,
  %``Extended phase space thermodynamics for charged and rotating black holes and Born-Infeld vacuum polarization,''
  JHEP {\bf 1211}, 110 (2012)
  [arXiv:1208.6251 [hep-th]];
  %%CITATION = ARXIV:1208.6251;%%
  %50 citations counted in INSPIRE as of 27 Aug 2014
S.~W.~Wei and Y.~X.~Liu,
  %``Critical phenomena and thermodynamic geometry of charged Gauss-Bonnet AdS black holes,''
  Phys.\ Rev.\ D {\bf 87}, no. 4, 044014 (2013)
  [arXiv:1209.1707 [gr-qc]];
  %%CITATION = ARXIV:1209.1707;%%
  %21 citations counted in INSPIRE as of 27 Aug 2014
A.~Belhaj, M.~Chabab, H.~El Moumni and M.~B.~Sedra,
  %``On Thermodynamics of AdS Black Holes in Arbitrary Dimensions,''
  Chin.\ Phys.\ Lett.\  {\bf 29}, 100401 (2012)
  [arXiv:1210.4617 [hep-th]];
  %%CITATION = ARXIV:1210.4617;%%
  %26 citations counted in INSPIRE as of 27 Aug 2014
 S.~H.~Hendi and M.~H.~Vahidinia,
  %``Extended phase space thermodynamics and P-V criticality of black holes with a nonlinear source,''
  Phys.\ Rev.\ D {\bf 88}, no. 8, 084045 (2013)
  [arXiv:1212.6128 [hep-th]];
  %%CITATION = ARXIV:1212.6128;%%
  %30 citations counted in INSPIRE as of 27 Aug 2014
 S.~Chen, X.~Liu, C.~Liu and J.~Jing,
  %``$P-V$ criticality of AdS black hole in $f®$ gravity,''
  Chin.\  Phys.\  Lett.\  {\bf 30}, 060401 (2013)
  [arXiv:1301.3234 [gr-qc]];
  %%CITATION = ARXIV:1301.3234;%%
  %13 citations counted in INSPIRE as of 27 Aug 2014
E.~Spallucci and A.~Smailagic,
  %``Maxwell's equal area law for charged Anti-deSitter black holes,''
  Phys.\ Lett.\ B {\bf 723}, 436 (2013)
  [arXiv:1305.3379 [hep-th]];
  %%CITATION = ARXIV:1305.3379;%%
  %20 citations counted in INSPIRE as of 27 Aug 2014
R.~Zhao, H.~H.~Zhao, M.~S.~Ma and L.~C.~Zhang,
  %``On the critical phenomena and thermodynamics of charged topological dilaton AdS black holes,''
  Eur.\ Phys.\ J.\ C {\bf 73}, 2645 (2013)
  [arXiv:1305.3725 [gr-qc]];
  %%CITATION = ARXIV:1305.3725;%%
  %19 citations counted in INSPIRE as of 27 Aug 2014
 A.~Belhaj, M.~Chabab, H.~El Moumni and M.~B.~Sedra,
  %``Critical Behaviors of 3D Black Holes with a Scalar Hair,''
  arXiv:1306.2518 [hep-th];
  %%CITATION = ARXIV:1306.2518;%%
  %26 citations counted in INSPIRE as of 27 Aug 2014
 N.~Altamirano, D.~Kubiznak and R.~B.~Mann,
  %``Reentrant phase transitions in rotating anti¨Cde Sitter black holes,''
  Phys.\ Rev.\ D {\bf 88}, no. 10, 101502 (2013)
  [arXiv:1306.5756 [hep-th]];
  %%CITATION = ARXIV:1306.5756;%%
  %24 citations counted in INSPIRE as of 27 Aug 2014
R.~G.~Cai, L.~M.~Cao, L.~Li and R.~Q.~Yang,
  %``P-V criticality in the extended phase space of Gauss-Bonnet black holes in AdS space,''
  JHEP {\bf 1309}, 005 (2013)
  [arXiv:1306.6233 [gr-qc]];
  %%CITATION = ARXIV:1306.6233;%%
  %27 citations counted in INSPIRE as of 27 Aug 2014
W.~Xu, H.~Xu and L.~Zhao,
  %``Gauss-Bonnet coupling constant as a free thermodynamical variable and the associated criticality,''
  Eur.\ Phys.\ J.\ C {\bf 74}, 2970 (2014)
  [arXiv:1311.3053 [gr-qc]];
  %%CITATION = ARXIV:1311.3053;%%
  %12 citations counted in INSPIRE as of 27 Aug 2014
J.~X.~Mo and W.~B.~Liu,
  %``Ehrenfest scheme for P-V criticality in the extended phase space of black holes,''
  Phys.\ Lett.\ B {\bf 727}, 336 (2013);
  %%CITATION = PHLTA,B727,336;%%
  %13 citations counted in INSPIRE as of 27 Aug 2014
D.~C.~Zou, S.~J.~Zhang and B.~Wang,
  %``Critical behavior of Born-Infeld AdS black holes in the extended phase space thermodynamics,''
  Phys.\ Rev.\ D {\bf 89}, 044002 (2014)
  [arXiv:1311.7299 [hep-th]];
  %%CITATION = ARXIV:1311.7299;%%
  %14 citations counted in INSPIRE as of 27 Aug 2014
J.~X.~Mo and W.~B.~Liu,
  %``$P-V$ criticality of topological black holes in Lovelock-Born-Infeld gravity,''
  Eur.\ Phys.\ J.\ C {\bf 74}, 2836 (2014)
  [arXiv:1401.0785 [gr-qc]];
  %%CITATION = ARXIV:1401.0785;%%
  %11 citations counted in INSPIRE as of 27 Aug 2014
 N.~Altamirano, D.~Kubiznak, R.~B.~Mann and Z.~Sherkatghanad,
  %``Thermodynamics of rotating black holes and black rings: phase transitions and thermodynamic volume,''
  Galaxies {\bf 2}, 89 (2014)
  [arXiv:1401.2586 [hep-th]];
  %%CITATION = ARXIV:1401.2586;%%
  %17 citations counted in INSPIRE as of 27 Aug 2014
 S.~W.~Wei and Y.~X.~Liu,
  %``Triple points and phase diagrams in the extended phase space of charged Gauss-Bonnet black holes in AdS space,''
  Phys.\ Rev.\ D {\bf 90}, 044057 (2014)
  [arXiv:1402.2837 [hep-th]];
  %%CITATION = ARXIV:1402.2837;%%
  %7 citations counted in INSPIRE as of 27 Aug 2014
 J.~X.~Mo,
  %``Ehrenfest scheme for the extended phase space of $f®$ black holes,''
  Europhys.\ Lett.\  {\bf 105}, 20003 (2014);
  %%CITATION = EULEE,105,20003;%%
  %4 citations counted in INSPIRE as of 27 Aug 2014
J.~X.~Mo, G.~Q.~Li and W.~B.~Liu,
  %``Another novel Ehrenfest scheme for P-V criticality of RN-AdS black holes,''
  Phys.\ Lett.\ B {\bf 730}, 111 (2014);
  %%CITATION = PHLTA,B730,111;%%
  %4 citations counted in INSPIRE as of 27 Aug 2014
J.~X.~Mo and W.~B.~Liu,
  %``Ehrenfest scheme for $P-V$ criticality of higher dimensional charged black holes, rotating black holes and Gauss-Bonnet AdS black holes,''
  Phys.\ Rev.\ D {\bf 89}, 084057 (2014)
  [arXiv:1404.3872 [gr-qc]];
  %%CITATION = ARXIV:1404.3872;%%
  %6 citations counted in INSPIRE as of 27 Aug 2014
  D.~C.~Zou, Y.~Liu and B.~Wang,
  %``Critical behavior of charged Gauss-Bonnet AdS black holes in the grand canonical ensemble,''
  arXiv:1404.5194 [hep-th];
  %%CITATION = ARXIV:1404.5194;%%
  %8 citations counted in INSPIRE as of 27 Aug 2014
 R.~Zhao, M.~Ma, H.~Zhao and L.~Zhang,
  %``The Critical Phenomena and Thermodynamics of the Reissner-Nordstrom-de Sitter Black Hole,''
  Adv.\ High Energy Phys.\  {\bf 2014}, 124854 (2014);
  %%CITATION = 00642,2014,124854;%%
  %1 citations counted in INSPIRE as of 27 Aug 2014
 H.~Xu, W.~Xu and L.~Zhao,
  %``Extended phase space thermodynamics for third order Lovelock black holes in diverse dimensions,''
  arXiv:1405.4143 [gr-qc];
  %%CITATION = ARXIV:1405.4143;%%
  %5 citations counted in INSPIRE as of 27 Aug 2014
 W.~Xu and L.~Zhao,
  %``Critical phenomena of static charged AdS black holes in conformal gravity,''
  Phys.\ Lett.\ B {\bf 736}, 214 (2014)
  [arXiv:1405.7665 [gr-qc]];
  %%CITATION = ARXIV:1405.7665;%%
  %3 citations counted in INSPIRE as of 27 Aug 2014
 M.~S.~Ma and Y.~Q.~Ma,
  %``Critical behaviors of black hole in an asymptotically safe gravity with cosmological constant,''
  arXiv:1405.7609 [hep-th];
  %%CITATION = ARXIV:1405.7609;%%
  %2 citations counted in INSPIRE as of 27 Aug 2014
 A.~M.~Frassino, D.~Kubiznak, R.~B.~Mann and F.~Simovic,
  %``Multiple Reentrant Phase Transitions and Triple Points in Lovelock Thermodynamics,''
  arXiv:1406.7015 [hep-th];
  %%CITATION = ARXIV:1406.7015;%%
  %3 citations counted in INSPIRE as of 27 Aug 2014
 G.~Q.~Li,
  %``Effects of dark energy on P-V criticality of charged AdS black holes,''
  Physics Letters B 735,256-260 (2014)
  [arXiv:1407.0011 [gr-qc]];
  %%CITATION = ARXIV:1407.0011;%%
  %1 citations counted in INSPIRE as of 27 Aug 2014
 C.~O.~Lee,
  %``The Extended Thermodynamic Properties of Taub-NUT/Bolt-AdS spaces,''
  arXiv:1408.2073 [hep-th];
  %%CITATION = ARXIV:1408.2073;%%
  %1 citations counted in INSPIRE as of 27 Aug 2014
 C.~V.~Johnson,
  %``The Extended Thermodynamic Phase Structure of Taub-NUT and Taub-Bolt,''
  arXiv:1406.4533 [hep-th].
  %%CITATION = ARXIV:1406.4533;%%
  %3 citations counted in INSPIRE as of 27 Aug 2014



%\bibitem{Dolan-1}B.P. Dolan, Class. Quantum Grav. {\bf31}, 035022(2014)[arXiv:1308.5403].

\bibitem{Dolan2014}B.~P.~Dolan,
  %``Bose condensation and branes,''
  arXiv:1406.7267 [hep-th].
  %%CITATION = ARXIV:1406.7267;%%
  %2 citations counted in INSPIRE as of 27 Aug 2014
 \bibitem{Johnson}  C.~V.~Johnson,
  %``Holographic Heat Engines,''
  arXiv:1404.5982 [hep-th].
  %%CITATION = ARXIV:1404.5982;%%
  %9 citations counted in INSPIRE as of 27 Aug 2014

%\cite{Kastor:2014dra}
\bibitem{Kastor:2014dra}
  D.~Kastor, S.~Ray and J.~Traschen,
  %``Chemical Potential in the First Law for Holographic Entanglement Entropy,''
  arXiv:1409.3521 [hep-th].
  %%CITATION = ARXIV:1409.3521;%%

\bibitem{Weinhold}F. Weinhold, J. Chem. Phys. {\bf63}, 2479 (1975).
\bibitem{Ruppeiner-2}G. Ruppeiner, Phys. Rev. A {\bf20}, 1608 (1979).

\bibitem{Rupreview}G.~Ruppeiner,
  %``Riemannian geometry in thermodynamic fluctuation theory,''
  Rev.\ Mod.\ Phys.\  {\bf 67}, 605 (1995)
  [Erratum-ibid.\  {\bf 68}, 313 (1996)].
  %%CITATION = RMPHA,67,605;%%
  %132 citations counted in INSPIRE as of 28 Aug 2014

\bibitem{Salamon}P. Salamon, J. D. Nulton and E. Ihrig, J. Chem. Phys.{\bf80},436(1984).
\bibitem{Berry} P. Salamon, E. Ihrig and R. S. Berry, J. Math. Phys.{\bf 24}, 2515(1983).
\bibitem{Mrugala} R. Mrugala, J. D. Nulton, J. C. Schon, and P. Salamon, Phys. Rev. A {\bf41}, 3156(1990).
\bibitem{Quevedo-1}H. Quevedo,  J. Math. Phys. {\bf48}, 013506(2007).
\bibitem{Quevedo-2} H. Quevedo, Gen. Relativ. Gravit. {\bf40} , 971 (2008).
\bibitem{Quevedo-3} H. Quevedo, A. Sanchez, S. Taj and A. Vazquez, Gen. Relativ.
Gravit. {\bf43}, 1153(2011).
\bibitem{Quevedo-4} A. Bravetti, D. Momeni, R. Myrzakulov and H. Quevedo, Gen. Relativ. Gravit.{\bf 45},1603(2013).

%\cite{Ferrara:1997tw}
\bibitem{Ferrara:1997tw}
  S.~Ferrara, G.~W.~Gibbons and R.~Kallosh,
  %``Black holes and critical points in moduli space,''
  Nucl.\ Phys.\ B {\bf 500}, 75 (1997)
  [hep-th/9702103].
  %%CITATION = HEP-TH/9702103;%%
  %385 citations counted in INSPIRE as of 09 Sep 2014


\bibitem{CC} R.~G.~Cai and J.~H.~Cho,
  %``Thermodynamic curvature of the BTZ black hole,''
  Phys.\ Rev.\ D {\bf 60}, 067502 (1999)
  [hep-th/9803261].
  %%CITATION = HEP-TH/9803261;%%
  %47 citations counted in INSPIRE as of 28 Aug 2014

 \bibitem{Aman:2003ug}
  J.~E.~Aman, I.~Bengtsson and N.~Pidokrajt,
  %``Geometry of black hole thermodynamics,''
  Gen.\ Rel.\ Grav.\  {\bf 35}, 1733 (2003)
  [gr-qc/0304015];
  %%CITATION = GR-QC/0304015;%%
  %81 citations counted in INSPIRE as of 12 Sep 2014
 %\cite{Aman:2005xk}
%\bibitem{Aman:2005xk}
  J.~E.~Aman and N.~Pidokrajt,
  %``Geometry of higher-dimensional black hole thermodynamics,''
  Phys.\ Rev.\ D {\bf 73}, 024017 (2006)
  [hep-th/0510139].
  %%CITATION = HEP-TH/0510139;%%
  %74 citations counted in INSPIRE as of 12 Sep 2014

\bibitem{Rup2013} G.~Ruppeiner,
  %``Thermodynamic curvature and black holes,''
  "Breaking of Supersymmetry and Ultraviolet Divergences in Extended
  Supergravity," Springer Proceedings in Physics Volume 153, 2014, pp 179-203
  [arXiv:1309.0901 [gr-qc]].
  %%CITATION = ARXIV:1309.0901;%%
  %3 citations counted in INSPIRE as of 28 Aug 2014

\bibitem{Mansoori:2013pna}
  S.~A.~H.~Mansoori and B.~Mirza,
  %``Correspondence of phase transition points and singularities of thermodynamic geometry of black holes,''
  Eur.\ Phys.\ J.\ C {\bf 74}, 2681 (2014)
  [arXiv:1308.1543 [gr-qc]]; S.~A.~H.~Mansoori, B.~Mirza and M.~
  Fazel,arXiv:1411.2582 [gr-qc].

\bibitem{Gubser}S.~S.~Gubser, I.~R.~Klebanov and A.~A.~Tseytlin,
  %``Coupling constant dependence in the thermodynamics of N=4 supersymmetric Yang-Mills theory,''
  Nucl.\ Phys.\ B {\bf 534}, 202 (1998)
  [hep-th/9805156].
  %%CITATION = HEP-TH/9805156;%%
  %307 citations counted in INSPIRE as of 28 Aug 2014


%\bibitem{myers1} A. Chamblin, R.Emparan, C. V. Johnson, R. C. Myers,Phys.Rev. D {\bf60},064018(1999).
%\bibitem{myers2} A. Chamblin, R.Emparan, C. V. Johnson, R. C. Myers,Phys.Rev.D  {\bf60},104026(1999).
%\bibitem{Maldacena}J. Maldacena, Adv. Theor. Math. Phys.{\bf 2}, 252(1998)[arXiv:9711200].
%\bibitem{Ruppeiner} G. Ruppeiner, Rev. Mod. Phys. {\bf67}, 605 (1995), {\bf68}, 313 (E) (1996).

\bibitem{Callen} H.B. Callen, {\it Thermodynamics and an Introduction to Thermostatistics}, second edition,
 1985, John Wiley and Sons.

\bibitem{Hendi}S.~H.~Hendi,
  %``Thermodynamic properties of asymptotically Reissner-Nordstr?m black holes,''
  Annals Phys.\  {\bf 346}, 42 (2014)
  [arXiv:1405.6996 [gr-qc]].
  %%CITATION = ARXIV:1405.6996;%%
  %1 citations counted in INSPIRE as of 12 Sep 2014

\bibitem{Suresh}J.~Suresh, R.~Tharanath, N.~Varghese and V.~C.~Kuriakose,
  %``The thermodynamics and thermodynamic geometry of the Park black hole,''
  Eur.\ Phys.\ J.\ C {\bf 74}, 2819 (2014)
  [arXiv:1403.4710 [gr-qc]].
  %%CITATION = ARXIV:1403.4710;%%
  %1 citations counted in INSPIRE as of 12 Sep 2014

\end{thebibliography}
\end{document}